\documentclass[prc]{revtex4}
\usepackage{graphicx}
\begin{document}
\title{A Co-moving Coordinate System for Relativistic Hydrodynamics}
\author{Scott Pratt}
\affiliation{Department of Physics and Astronomy,
Michigan State University\\
East Lansing, Michigan 48824-1321}
\date{\today}

 
\begin{abstract}
The equations of relativistic hydrodynamics are transformed so that steps forward in time preserves local simultaneity. In these variables, the space-time coordinates of neighboring points on the mesh are simultaneous according to co-moving observers. Aside from the time step varying as a function of the location on the mesh, the local velocity gradient and the local density then evolve according to non-relativistic equations of motion. Analytic solutions are found for two one-dimensional cases with constant speed of sound. One solution has a Gaussian density profile when mapped into the new coordinates. That solution is analyzed for the effects of longitudinal acceleration in relativistic heavy ion collisions at RHIC, especially in regards to two-particle correlation measurements of the longitudinal size.

\end{abstract}

\maketitle

\section{Introduction and Theory}
\label{sec:intro}

The large collective flow observed at the Relativistic Heavy Ion Collider (RHIC) have revitalized interest in relativistic hydrodynamics for nuclear collisions \cite{Huovinen:2003fa,Kolb:2003dz,Teaney:2001av,Rischke:1996nq}. Numerical approaches for solving non-relativistic equations can be divided into numerous categories. Information regarding the densities and collective velocities might be stored for static mesh points (Eulerian), or for mesh points that move with the fluid (Lagrangian). Other approaches discard the notion of a mesh and instead follow the motion of particles which represent fluid elements \cite{smoothhydro,smoothhydrobook,Aguiar:2000hw}, thus making the solution similar to that of molecular dynamics. Adding viscosity and heat conduction brings about an even wider range of approaches. For instance, once heat diffusion is introduced one must choose whether the velocity refers to that of the energy \cite{landauframe} (Landau frame) or that of the number density \cite{eckartframe} (Eckart frame), or even of the entropy density \cite{carterframe}. In addition to specifying the character of the mesh and the equations, there exist innumerable algorithms.

Relativistic modeling involves another class of choices, as there may be good reasons to consider meshes where information is not stored for a space-time hyper-surface defined by a fixed Cartesian time, but rather for some alternately  defined hyper-surfaces. For instance, cosmological modeling might use the proper time, i.e., the time as measured by a local observer. For relativistic collision, where there is semi-Hubble-like flow along the $z$-axis, one might use the Hwo-Bjorken time \cite{Hwa:1974gn,Bjorken:1982qr}, $\tau=\sqrt{t^2-z^2}$, which is the time as measured by an observer moving with constant velocity $v=z/t$. If the matter accelerates along the $z$ axis, this time would differ from the time measured by an observer moving with the fluid. Once the fluid velocities significantly differ from that of the mesh, relativistic effects become important.

In this study we investigate the prospects of propagating so that after each point $i$ moves forward by $\delta\tau_i$, the hyper-surface will satisfy local simultaneity, 
\begin{equation}
u\cdot dr=0,
\end{equation}
where $dr^\mu$ is the differential distance between neighboring points on the hyper-surface, and $u$ is the four-velocity, where in the frame of fluid $u=(1,0,0,0)$. Thus, in the frame of the fluid the time difference $dt$ between neighboring mesh points is zero. After presenting the equations of motion that preserve local simultaneity in the next section, we present analytic solutions in these coordinates in the next two solutions. The first solution is for a Gaussian-shaped entropy distribution that is initially at rest, while the second is a restatement of the Khalatnikov-Landau solution for matter leaving a static shock \cite{khalatnikov,landau}. Both solutions are for the particular case of purely one-dimensional motion with a constant speed of sound. Sec. \ref{sec:comparison} explores the manifestations of longitudinal acceleration and compares the analytic Gaussian case with a numerical solution for the case with initial collective flow given by the Hwa-Bjorken velocity profile, $u_z=z/\tau$. In particular, the degree to which the velocity gradient differs from $1/\tau$ is discussed for its importance in affecting conclusions about lifetime from two-particle correlations. The prospects and limitations for these approaches are reviewed in Sec. \ref{sec:summary}, after which an appendix is included reviewing other analytic solutions to relativistic hydrodynamics.

\section{Maintaining Simultaneity in Relativistic Hydrodynamics}
\label{sec:generaltheory}

Here, we express the equations for relativistic hydrodynamics in terms of coordinates relevant for a set of observers traveling with the fluid. Each observer initializes his clock so that, in each observer's frame, evolution will begin simultaneously for the neighboring observers. Such is the case for the commonly used Hwa-Bjorken assumption \cite{Hwa:1974gn,Bjorken:1982qr}, where the collective velocity is given by $u^z=z/\tau$, with $\tau$ being the proper time, i.e., the time measured by each observer in their frame. In the Hwa-Bjorken picture, there is no acceleration and local simultaneity ($u\cdot x=0$) is preserved if each co-moving observer propagates forward equal amounts in proper time. However, as will be shown here, simultaneity will not be maintained in the presence of acceleration. For accelerating cases, each co-moving observer must propagate forward by different time steps. 

The equations of motion for a relativistic hydrodynamic fluid conserving entropy are:
\begin{eqnarray}
\label{eq:motionaa}
(u\cdot\partial) s &=& -s \partial\cdot u\\
(u\cdot\partial) u^\alpha 
&=& \frac{1}{P+\epsilon} (\partial^\alpha-u^\alpha u\cdot\partial)P,
\end{eqnarray}
where $s$ is the entropy density as measured by a co-mover. Throughout this section, we ignore the general case where other currents are conserved, thus making $P$ a function of $s$ only. For a specific differential fluid element, an observer moving in the frame of the fluid observes a change in the velocity and entropy density,
\begin{eqnarray}
\label{eq:motiona}
\delta {\bf v}&=&-\frac{\nabla P}{P+\epsilon}\delta\tau,\\
\delta s&=&-s(\nabla\cdot{\bf v})\delta\tau
\end{eqnarray}
where $\delta\tau$ is the differential time step and ${\bf r}$ and ${\bf v}$ are the position and velocity measured by an observer moving with the same velocity as the matter.

The equations of motion can be integrated forward numerically on a Lagrangian mesh, one that moves with the fluid, by solving for the time development of the separation between neighboring mesh points, $\Delta{\bf r}$, and the relative velocity of the two points, $\Delta {\bf v}$, where $\Delta{\bf r}$ and $\Delta{\bf v}$ are always measured by an observer traveling with the fluid between the two mesh points. Since the velocities are, by definition, small according to this observer, the motion appears non-relativistic. After solving for the non-relativistic evolution of $\Delta{\bf v}$ and $\Delta{\bf r}$, one could perform incremental relativistic velocity additions and Lorentz transformations to solve for the relativistic evolution. Thus, it would appear that relativistic complications could be avoided until after the non-relativistic evolution of the mesh is determined.

The algorithm described above assumes that $\Delta{\bf r}$ represents the separation of two simultaneous space-time points, i.e., $\Delta t=0$ in the frame of the fluid. Unfortunately, even if the mesh is originally defined with neighboring points being simultaneous, the simultaneity will be destroyed after propagating each point forward a fixed amount of proper time if the points undergo acceleration. Here, we show that simultaneity can be maintained if different points on the hyper-surface are propagated forward by different steps in proper time.

Simultaneity between two neighboring points $x_1$ and $x_2$ with velocities $u_1$ and $u_2$ can be stated by the relation,
\begin{equation}
(u_1+u_2)\cdot(x_2-x_1)=0,
\end{equation}
where $\Delta x=x_2-x_1$ is small and $(1/2)(u_1+u_2)$ is the four velocity of a co-moving observer between the two points of the fluid. Maintaining simultaneity requires:
\begin{equation}
\label{eq:simula}
(u_1+u_2+a\delta\tau_1+a\delta\tau_2)\cdot(x_1-x_2+u_1\delta\tau_1-u_2\delta\tau_2)=0,
\end{equation}
where $a^\mu$ is the acceleration. Since accelerations have no time components in the frame of the matter, $a\cdot u=0$, and since $(u_1-u_2)\cdot(u_1+u_2)$ is also zero, one can rewrite Eq. (\ref{eq:simula}) to find:
\begin{eqnarray}
\label{eq:simulb}
\left[(u_1+u_2)+a(\delta\tau_1+\delta\tau_2)\right]
\cdot\left[\Delta x+(u_1-u_2)(\delta\tau_1+\delta\tau_2)/2
-(u_1+u_2)(\delta\tau_2-\delta\tau_1)/2\right]&=&0,\\
\nonumber
(\delta\tau_1+\delta\tau_2) a\cdot(\Delta x)-2(\delta\tau_2-\delta\tau_1)&=&0,\\
\nonumber
\delta\tau_2=\delta\tau_1(1-a\cdot\Delta x),
\end{eqnarray}
where it has also been assumed that $\Delta x$ and $\delta\tau$ are small, i.e., $(u_1+u_2)\cdot(u_1+u_2)\approx 4$. The last expression in Eq.s (\ref{eq:simulb}) can be integrated to find an expression for ratio of time steps between an observer at $x$ and a reference observer at $x_R$,
\begin{eqnarray}
\label{eq:alphadef}
\delta\tau&=&\alpha\delta\tau_R,\\
\nonumber
\alpha(x)&=&\exp\left\{-\int_{x_R}^{x} a\cdot dx' \right\}\\
\nonumber
&=&\exp\left\{\int c^2 d\ln s \right\},
\end{eqnarray}
where the last step used the definition of the speed of sound for the case where $P$ is expressed as a function of the entropy density,  
\begin{equation}
c_s^2=s\frac{dP/ds}{P+\epsilon}
\end{equation}

Thus, although the equations of motion, Eq.s (\ref{eq:motiona}), appear simple to solve by working in a co-moving frame where distance is measured along a path satisfying $u\cdot dx=0$, the equations can not be integrated forward by fixed $\delta\tau$ without violating the condition of simultaneity. Instead, one should integrate forward in time by an amount, $d\tau =\alpha(x) d\tau_R$, where $\tau_R$ refers to the time measured by a reference observer at $x_R$.

The equations of motion can be rewritten in terms of $d\tau_R$ rather than $d\tau$,
\begin{eqnarray}
\label{eq:motionb}
\delta s 
&=& -\alpha s (\nabla\cdot{\bf v}) \delta\tau_R,\\
\nonumber
\delta {\bf v}
&=& -\frac{\alpha}{P+\epsilon} (\nabla P)
\delta\tau_R.\end{eqnarray}
These expressions differ from the non-relativistic equations of motion due to the presence of $\alpha$ on the right-hand sides of both equations. These equations are especially well suited for Eulerian approaches, where the mesh moves with the fluid, and one need only solve the equations of motion for the evolution of differences of position and velocity between neighboring points, as measured by a co-moving observer. 

Solution in more than one dimension of Eq. (\ref{eq:motionb}) requires that the flow is ir-rotational. For rotational flow, the definition of simultaneity, $u\cdot dx=0$, can not uniquely map to a hyper-surface extending over all space. For instance, one can consider a ring of particles in uniform circular motion of a radius $R$ at a velocity $v$. If two neighboring particles would synchronize their clocks as to be simultaneous in the frame of an observer moving with the particles, the laboratory observer would view this synchronization event as happening at two different times separated by $\delta t=v\Delta x$. If the synchronization were to be performed clockwise around the ring, the times would be off by a factor $2\pi Rv$ after returning to the original point. The problem with rotational flow is also encountered when considering for $\alpha=\exp(-\int dx\cdot a)$ which is only path independent if $\nabla\times \vec{a}=0$. If $\nabla\times \vec{v}=0$ everywhere for the initial condition, and if $\nabla\times\vec{a}=0$ for all times, the flow will remain ir-rotational. The requirement that $\nabla\times\vec{a}=0$ is maintained when the pressure depends solely on the energy density, but can be violated if the pressure also depends on a particle density $n$. This can be shown by considering
\begin{eqnarray}
\nabla\times\vec{a}&=&\nabla\times\frac{\nabla P(\epsilon,n)}{P(\epsilon,n)+\epsilon}\\
&=&-\frac{\nabla\epsilon\times\nabla P}{(P+\epsilon)^2}\\
&=&-\frac{\partial P}{\partial n}\frac{\nabla \epsilon\times\nabla n}{(P+\epsilon)^2}.
\end{eqnarray}
For the physics of RHIC, and especially for the physics of the LHC, the particle densities are small and rotational aspects of initial flow should be small. The flow would indeed be rotational at non-zero impact parameters, even at midrapidity \cite{Lisa:2000ip}. However, these effects should be less pronounced at higher beam energies if the production mechanism becomes increasingly boost-invariant.

\section{One-Dimensional Gaussian Solution}
\label{sec:gaussian}

Here, we present an exact analytic solution to the hydrodynamic equations of motion in Eq. (\ref{eq:motionb}) for the specific case of one-dimensional motion where the speed of sound is constant. The solution is found by first mapping the coordinate $\ell$, which measures the distance along a space-time path of simultaneity, to a new coordinate $\tilde{\ell}$. The transformation to the new coordinate $\tilde{\ell}$ will be chosen to cancel the factor of $\alpha$ in the equations of motion.
\begin{equation}
d\tilde{\ell}=\alpha d\ell,~~~~\tilde{\ell}=\int_0^\ell d\ell'\alpha(\ell').
\end{equation}
In these coordinates, the equations of motion become
\begin{eqnarray}
\delta s 
&=& -s \frac{\partial v}{\partial\tilde{\ell}} ~\delta\tau_R,\\
\nonumber
\delta v
&=& -\frac{1}{P+\epsilon}\frac{\partial P}{\partial\tilde{\ell}}~\delta\tau_R.
\end{eqnarray}
Expanding $\delta s$ and $\delta v$,
\begin{eqnarray}
\frac{\partial s}{\partial\tau_R}+\frac{\partial s}{\partial\tilde{\ell}}\frac{d\tilde{\ell}}{d\tau_R}&=&-s \frac{\partial v}{\partial\tilde{\ell}},\\
\nonumber
\frac{\partial y}{\partial\tau_R}+\frac{\partial y}{\partial\tilde{\ell}}\frac{d\tilde{\ell}}{d\tau_R}&=&-\frac{1}{P+\epsilon}\frac{\partial P}{\partial\tilde{\ell}}.
\end{eqnarray}
Here, we have identified the sum of small changes of the velocity as measured by co-movers, $\Delta v$, as the rapidity $y$. These would look exactly like the non-relativistic expressions if the velocity could be identified with $d\tilde{\ell}/d\tau_R$. This will not always be satisfied. In general, one can write $\tilde{\ell}$ as a sum over contributions between neighboring cells of length $\Delta\ell$,
\begin{eqnarray}
\frac{d\tilde{\ell}}{d\tau_R}&=&\frac{d}{d\tau_R}\sum_i \Delta \ell_i
\alpha(\ell_i)\\
\nonumber
&=&\sum_i \alpha_i\frac{d\Delta\ell}{d\tau_R}+\sum_i \Delta\ell_i\frac{d\alpha_i}{d\tau_R}\\
\nonumber
&=&\sum_i \frac{d\Delta\ell}{d\tau}+\sum_i \Delta\ell_i\frac{d\alpha_i}{d\tau_R}\\
\nonumber
&=&y+\int \sum_i\frac{d\alpha_i}{d\tau_R}.
\end{eqnarray}
The last term will vanish only if the co-moving time derivate of $\alpha$ vanishes. This criteria will be satisfied for our specific case where the speed of sound is constant and the entropy profile is set by a scaling function. From the definition of $\alpha$ in Eq. (\ref{eq:alphadef}) one can see that for a constant speed of sound,
\begin{equation}
\alpha=\exp\left\{c_s^2 \ln(s/s_R)\right\}=\left(\frac{s}{s_R}\right)^{c_s^2}.
\end{equation}
For a scaling solution, $s(\tilde{\ell},\tau_R)=F(\tilde{\ell}/R(\tau_R))$, and the shape of the entropy profile is unchanged through time, aside from changing the scale $R$. The ratio $s/s_R$ will be then stay constant for each co-mover. 

Here, we present a solution where the entropy profile is assumed to follow a Gaussian profile,
\begin{equation}
s(\tilde{\ell},\tau_R)=\frac{1}{R(\tau_R)}
\exp\left\{-\frac{\tilde{\ell}^2}{2R(\tau_R)^2}\right\}.
\end{equation}
Furthermore, the velocity profile is assumed to be linear in $\tilde{\ell}$,
\begin{equation}
\label{eq:yofell}
\frac{y}{c_s}=A(\tilde{t}\equiv c_s\tau_R/R_0)\frac{\tilde{\ell}}{R_0}.
\end{equation}
Since the entropy profile scales with time, the equations of motion are those of non-relativistic hydrodynamics,
\begin{eqnarray}
\frac{\partial s}{\partial \tau_R}+\frac{\partial (sy)}{\partial \tilde{\ell}}&=&0\\
\nonumber
\frac{\partial y}{\partial \tau_R}+y\frac{\partial y}{\partial \tilde{\ell}}&=&-c_s^2\frac{\partial \ln s}{\partial \tilde{\ell}}.
\end{eqnarray}
After applying the hydrodynamic equations of motion to the assumed forms for $s$ and $y$, the dependence on $\tilde{\ell}$ factors away and leaves simple differential equations of motion for $A(\tilde{t}\equiv c_s\tau_R/R_0)$ and for $r(\tilde{t})\equiv R(\tilde{t})/R_0$,
\begin{eqnarray}
\label{eq:tauR}
\frac{dA}{d\tilde{t}}+A^2&=&\frac{1}{r^2},\\
\nonumber
A&=&\frac{dr/dt}{r}.
\end{eqnarray}
Substituting for $A$ provides a differential equation for $r$,
\begin{equation}
\frac{d^2r}{d\tilde{t}^2}=\frac{1}{r},
\end{equation}
which can be solved by making an analogy with classical mechanics where one wishes to find the trajectory for a force behaving as $1/r$. The solution is
\begin{eqnarray}
\label{eq:tofr}
\tilde{t}&=&\int_1^r \frac{dx}{\sqrt{2\ln(x)}}=2^{1/2}\int_0^{\sqrt{\ln(r)}} dy~ e^{y^2},\\
\nonumber
&=&2^{1/2}rD(\sqrt{\ln r}),
\end{eqnarray}
where $D(x)$ is Dawson's integral function \cite{abramowitzstegunpage298} defined by the integral,
\begin{equation}
D(x)\equiv e^{-x^2}\int_0^x du~e^{u^2}.
\end{equation}
$D(x)$ rises proportional to $x$ for low $x$ and falls as $1/2x$ for large $x$. Here, $t=0$ refers to the time at which the velocity gradient vanishes, and $r$ represents the ratio of the $R$ to the radius at that time. The expression is analytic, but in practice, evaluating the expression might require programming the function, which diminishes the glamour of finding an analytic expression. This expression must be inverted to find $r$ as a function of $\tilde{t}$. The velocity gradient can then be expressed in terms of $r$,
\begin{equation}
\label{eq:Asolution}
A(\tilde{t})=\frac{1}{r(\tilde{t})}\sqrt{\ln(r^2)}
\end{equation}

A disquieting aspect of the solution is that the entropy density is $s=dS/d\ell$, while the solution conserves the integral $\int sd\tilde{\ell}$. However, one can understand that both $dS/d\tilde{\ell}$ and $dS/d\ell$ represent conserved charges by considering a co-moving cell. The two quantities differ by a factor $\alpha=d\tilde{\ell}/d\ell$, and since $\alpha$ remains constant and since the entropy is conserved within each cell individually, both $\int sd\ell$ and $\int sd\tilde{\ell}$ survive as constants of the motion.

If one solves the equations of motion in terms of $\tau_R$ and $\tilde{\ell}$, the state of the system in terms of the coordinates determined in a fixed reference frame, $x$ and $t$, requires inversion of the mapping procedure. To find $s$ and $u$ for a given time and position in the $x$ direction, one could follow the following steps:\\
\noindent
(i) As a function of $\tilde{\ell}$ and $\tau_R$, use the equations of motion to solve for the entropy density, $s$, the rapidity, $v$, and the factor $\alpha$.

\noindent
(ii) Begin with $\tau_R=t$ and $\tilde{\ell}=0$

\noindent
(iii) For small $\Delta x$, choose $\Delta\tau$ and $\Delta\ell$ to satisfy the conditions,
\begin{eqnarray}
\Delta x&=& \cosh(y)\Delta\ell+\sinh(y)\Delta\tau,\\
\nonumber
0&=&\cosh(y)\Delta\tau+\sinh(y)\Delta\ell.
\end{eqnarray}

\noindent
(iv) Find $\Delta\tilde{\ell}=\alpha\Delta\ell$ and $\Delta\tau_R=\alpha\Delta\tau$. Increment $\tau_R$ and $\tilde{\ell}$.

\noindent
(v) Repeat steps (iii) - (iv) until the point $(t,x)$ is reached.

\noindent
Solutions for different values of $R_0$ and $c_s$ can be transformed into one another by scaling $\tau_R$ and $R$ through the dimensionless variables $\tilde{t}$ and $r$ described above. However, this simple equivalence is lost when one maps to the coordinates in fixed reference frame, $x$ and $t$. The mapping procedure will then depend on the ratio of the speed of sound to the speed of light.

\begin{figure}
\centerline{\includegraphics[width=3.0in]{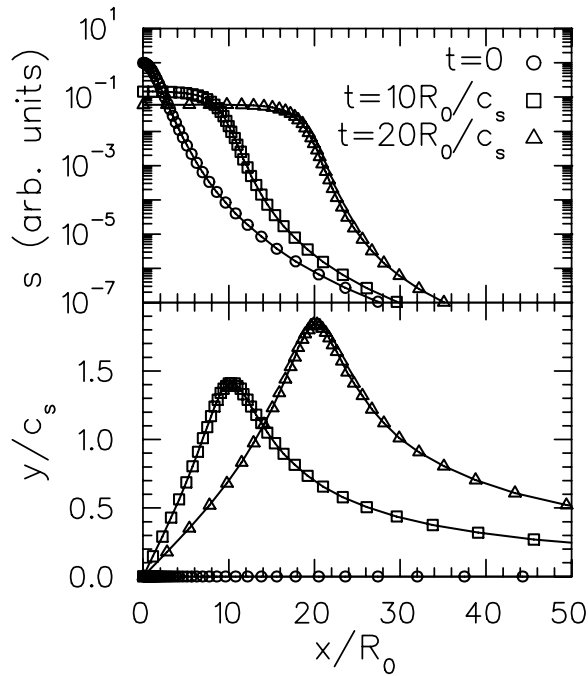}}
\caption{\label{fig:anal}
The analytic solution is shown after being transformed into Cartesian space-time coordinates for three times, $t=0$ (circles), $t=10R_0/c_s$ (squares) and $t=20R_0/c_s$ (triangles). The initial entropy profile is Gaussian in the variable $\tilde{\ell}$ but in Cartesian coordinates behaves like a power-law in the tail. The velocity profiles turn over at large $x$ is due to the falling slope of the entropy profile combined with the effects of transforming the solution into Cartesian coordinates. }
\end{figure}

Figure \ref{fig:anal} shows the entropy density and collective rapidity as a function of the position and time in a Cartesian reference frame for the case where the speed of sound is $c_s^2=0.2$. In terms of $\tilde{\ell}$, the profile is of a Gaussian form, but after scaling to $d\ell=d\tilde{\ell}/\alpha$, then performing the Lorentz transformations to Cartesian space-time coordinates, the profile is decidedly non-Gaussian. This is evident for the entropy density, $s$, plotted in the upper panel of Fig. \ref{fig:anal} for $t=0$. On a logarithmic scale, a Gaussian is concave downward for all $x$; whereas the scaled Gaussian picks up a power-law tail, which after some non-trivial expansions can be shown to have the form,
\begin{equation}
s(x\rightarrow\infty,t=0)\approx
\left(\frac{u}{2\sqrt{\ln u}}\right)^{-1/c^2},~u\equiv c_sx/(R_0\sqrt{2}).
\end{equation}
Since the acceleration for a fixed speed of sound is proportional to the logarithmic derivative of the density, the reversal of the concavity in $s(x)$ results in an acceleration that is maximized for intermediate values as illustrated by the plot of rapidity vs. position for subsequent times displayed in the lower panel of Fig. \ref{fig:anal}. This is distinctly unlike the behavior one would observe for a non-relativistic system \cite{Csanad:2003qa}, where the acceleration rises linearly at all $x$.

The solutions illustrated in Fig. \ref{fig:anal} were also checked for conservation of energy and entropy.
\begin{equation}
E_{\rm tot}(t)=\int dx T^{00}(x,t),~~~S_{\rm tot}=\int dx~u^0(x,t)s(x,t),
\end{equation}
where the stress-energy tensor was calculated from $s(x,t)$ and the collective four velocity $u$,
\begin{eqnarray}
T^{\alpha\beta}(x,t)&=&(P+\epsilon)u^\alpha u^\beta -Pg^{\alpha\beta},\\
\nonumber
P&=&s^{1+c_s^2},~~~\epsilon=P/c_s^2.
\end{eqnarray}
Although the expressions used to calculate $u$ and $s$ are analytic, calculation of the Dawson's integrals, and the mapping to a fixed reference frame involved numerical integrations which introduced small errors. These errors were less than 0.1\%, and fell with increasing granularity in the calculations. Furthermore, the net momentum of all matter moving to the right was tested for consistency with the integrated impulse,
\begin{equation}
\int dx~T^{0x}(x,t)=\int_0^t dt' P(x=0,t').
\end{equation}
This condition was also satisfied to high precision.

Despite the fact that the space-time evolution of the entropy density is rather complicated, the entropy per rapidity has a simple form. To see this, we begin with the fact that $s=dS/d\ell$, then apply the chain rule to find $dS/dy$,
\begin{equation}
\frac{dS}{dy}=s\frac{d\ell}{d\tilde{\ell}}\frac{d\tilde{\ell}}{dy}
=\frac{1}{R(\tau_R)}
\exp\left\{-\frac{\tilde{\ell}^2}{2R(\tau_R)^2}\right\}
\frac{d\ell}{d\tilde{\ell}}\frac{d\tilde{\ell}}{dy}.
\end{equation}
After using the simple linear form, $y=A\tilde{\ell}/R_0$, in Eq. (\ref{eq:yofell}), and integrating Eq. (\ref{eq:alphadef}) to find, 
\begin{equation}
\alpha=d\tilde{\ell}/d\ell=e^{-c_s^2\tilde{\ell}^2/(2R^2)}.
\end{equation}
After substituting for $\tilde{\ell}$ using Eq. (\ref{eq:yofell}) the rapidity distribution becomes
\begin{eqnarray}
\label{eq:dsdygauss}
\frac{dS}{dy}&=&\frac{1}{\sqrt{\ln(r^2)}}\exp\left\{\frac{-(1-c_s^2)y^2}{2A^2c_s^2r^2}\right\}\\
&=&\frac{1}{\sqrt{\ln(r^2)}}\exp\left\{\frac{-(1-c_s^2)y^2}{2c_s^2\ln(r^2)}\right\}
\end{eqnarray}
Here, we have also substituted for $rA=[\ln(r^2)]^{1/2}$ using Eq. (\ref{eq:Asolution}), and $r$ is again related to $R$ by scaling, $r=R/R_0$. The Gaussian width of the rapidty distribution is thus,
\begin{equation}
\label{eq:sigmay}
\sigma_y=c_s\sqrt{\frac{\ln (r^2)}{1-c_s^2}}.
\end{equation}
For large times, the width of the rapidity distribution in Eq. (\ref{eq:dsdygauss}), grows logarithmically with time as can be seen by considering the large $r$ behavior of the Dawson's integral function in Eq. (\ref{eq:tofr}), $\sqrt{\ln(r)}\sim \sqrt{\ln(t)}$. We emphasize that this simple Gaussian form relies on the assumption that the breakup is sudden along a hyper-surface of simultaneity.

Experimentally, the breakup time is inferred from two-particle correlations through the measurement of $R_{\rm long}$ which represents the size along the beam axis of the phase space distribution of particles with zero rapidity. If the collective motion covers large amounts of rapidity, the size of the phase space region is defined by the point at which the collective velocity, $R_{\rm long}dv/dz$, surpasses the thermal velocity \cite{Pratt:1986cc,Makhlin:1987gm}.
\begin{equation}
R_{\rm long}=\frac{v_{\rm therm}}{dv/dz}.
\end{equation}
For the Hwa-Bjorken solution, where there is no longitudinal acceleration, the collective velocity is always $z/t$. The velocity gradient is then $1/\tau$, where $\tau$ is the proper time elapsed since the beginning of the collision. In the no-acceleration Hwa-Bjorken ansatz, the time $\tau$ is then inferred by 
\begin{equation}
\tau=\frac{R_{\rm long}}{v_{\rm therm}}.
\end{equation}
Here, $R_{\rm long}$ is experimentally determined and the thermal velocity is inferred from spectra or blast-wave models. The simple statement above is modified by collective transverse flow, though the basic linear nature of the dependencies is preserved.

The presence of longitudinal acceleration would alter the inference of $\tau$ by voiding the Hwa-Bjorken equivalence $dv/dz=1/\tau$. In the analytic Gaussian solution shown here, the velocity gradient is given by Eq. (\ref{eq:Asolution}),
\begin{eqnarray}
\label{eq:dvdzanal}
\frac{dv}{dz}&=&\frac{c_s}{R}\sqrt{2\ln(R/R_0)}\\
\nonumber
&=&\frac{\sigma_y}{R}\sqrt{1-c_s^2}\\
\nonumber
&=&\frac{1}{\tau}\left[2xD(x)\right],~x\equiv \sigma_y\sqrt{(1-c_s^2)/2c_s^2}.
\end{eqnarray}
Here, the second line used the relation for the rapidity width, Eq. (\ref{eq:sigmay}), and the final step used Eq. (\ref{eq:tofr}). 

The quantity in the square brackets in Eq. (\ref{eq:dvdzanal}) now represents the correction to the Bjorken formula for the lifetime,
\begin{equation}
\label{eq:taudistortion}
\tau=\frac{R_{\rm long}}{v_{\rm therm}}\left[2xD(x)\right].
\end{equation}
For large rapidty widths, the Dawson's integral function can be expanded for large $x$,
\begin{equation}
D(x\rightarrow\infty)=\frac{1}{2x}+\frac{1}{4x^3}+\frac{3}{8x^5}
+\cdots \frac{(2n-1)!!}{2^{(n+1)}x^{(2n+1)}}+\cdots
\end{equation}
The first order term reproduces the Hwa-Bjorken result. At RHIC, the variance of the pion rapidity distribution is a bit over 2.0 \cite{Back:2005hs}. However, here $y$ refers to the source rapidity which should combine in quadrature with the thermal spread, $\sim 1$ to result in the pion rapidity spread. Thus, $\sigma_y$ should be $\sim 1.7$ here, which if we combine this value of $\sigma$ with a speed of sound,  $c_s^2\sim 0.20$, $x$ should in the neighborhood of 2.4, which would suggest that the Hwa-Bjorken estimate underestimates the lifetime by approximately 10\%, which is similar in strength to what was seen by comparing three-dimensional to boost-invariant numerical hydrodynamics by Renk, \cite{Renk:2005ws}. 

\section{Comparison with Hwa-Bjorken Initial Condition}
\label{sec:comparison}

The analytic solution of the previous section was predicated on the physical picture of complete stopping. Experiment strongly suggests, mainly through baryon distributions, that there is significant stopping, which invalidates the analytic expressions derived in the previous section. However, it is often the case that for large times solutions approach analytic forms even though the initial conditions differed significantly from the analytic form. To investigate this possibility, we compare the solutions for the same speed of sound, $c_s^2=0.2$, but for two very different initial conditions:
\begin{itemize}
\item[(a)] An analytic solution which assumes complete stopping at $\tau=0$. An initial Gaussian size of 0.06416 fm was chosen so that when the system expanded to its breakup time of $10$ fm/$c$, the rapidity width would match that estimated for the final state as discussed in the previous section, $\sigma_y(\tau=10 {\rm fm}/c)=1.7$ \cite{Heinz:2004et}. Here, the variance is defined using the entropy density as a weight,
\begin{equation}
\sigma_y^2(\tau)=\frac{\int d\ell~y^2s(\ell,\tau)}
{\int d\ell~s(\ell,\tau)}
\end{equation}
\item[(b)] Numerical solutions of hydrodynamics assuming a Hwa-Bjorken ansatz for the initial conditions, with the rapidity at time $\tau_0=0.5$ fm/$c$ given by,
\begin{equation}
y=\frac{\ell}{\tau_0},
\end{equation}
where $\ell$ is defined in the same way as in Sec. \ref{sec:generaltheory} as the net distance separating co-movers. The initial entropy density is chosen to be Gaussian,
\begin{equation}
\frac{dS}{d\ell}\propto e^{-\ell^2/2R_0^2}.
\end{equation}
These initial conditions differ from that used in the analytic solution both in that the matter is not moving, and in that the Gaussian form was applied to $\ell$, rather than the scaled variable $\tilde{\ell}$. The initial time $\tau_0$ was chosen rather arbitrarily to be 0.5 fm/$c$, and the initial size $R_0$ was then adjusted so that the variance $\sigma_y$ would again be 1.7 at a breakup time $\tau=10$ fm/c.  The size required to achieve this fit is $R_0=0.55$ fm.
\end{itemize}

\begin{figure}[hbt]
\centerline{\includegraphics[width=0.5\textwidth]{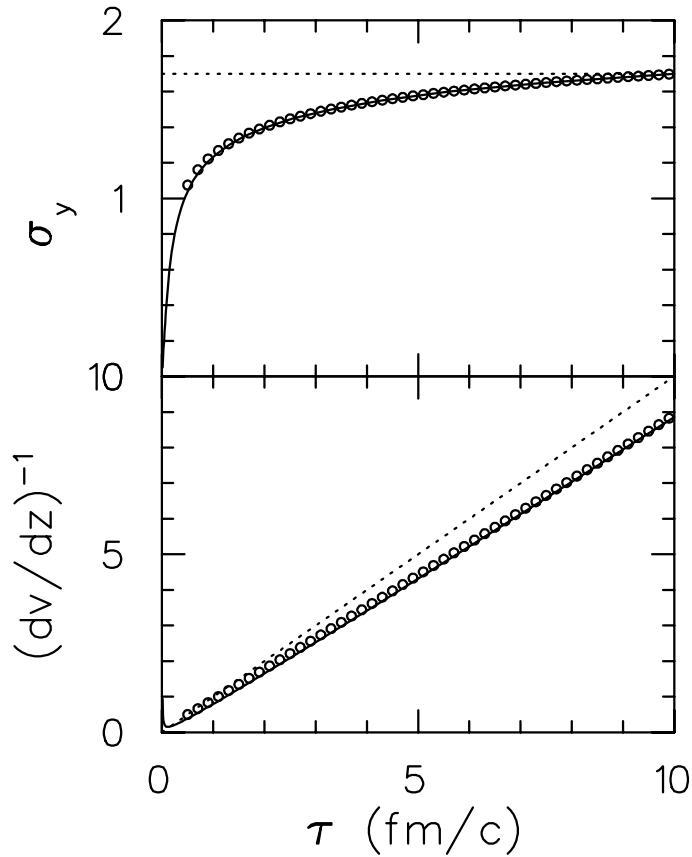}}
\caption{\label{fig:comparison}
The rapidity width and inverse velocity gradient are shown as a function of $\tau$ for the analytic solution (solid lines) and a numerical solution of a system whose initial collective flow at $\tau_0=0.5$ fm/$c$ was that of a Hwa-Bjorken model. For the analytic solution, the matter was initially stopped with a size chosen so that the rapidity width would grow to 1.7 by 10 fm/$c$. For the numerical solution, the initial width of the rapidity distribution was chosen so that it would also reach $\sigma_y=1.7$ at $\tau=10$ fm/$c$. Even though the initial velocity gradient is zero in the analytic solution, the models behave similarly for all $\tau>\tau_0$. The dashed lines show what would be expected in a Hwa-Bjorken expansion that neglects acceleration. The upper panel shows that acceleration increases the rapidity width by over 50\%. In the lower panel, the difference between the calculations with acceleration and the acceleration-less solution demonstrates the error associated with assigning the inverse velocity gradient, which is inferred from correlation measurements, with the lifetime. Accounting for acceleration should increase the estimate of the breakup time by $\sim$ 10\%.}
\end{figure}

\begin{figure}[hbt]
\centerline{\includegraphics[width=0.5\textwidth]{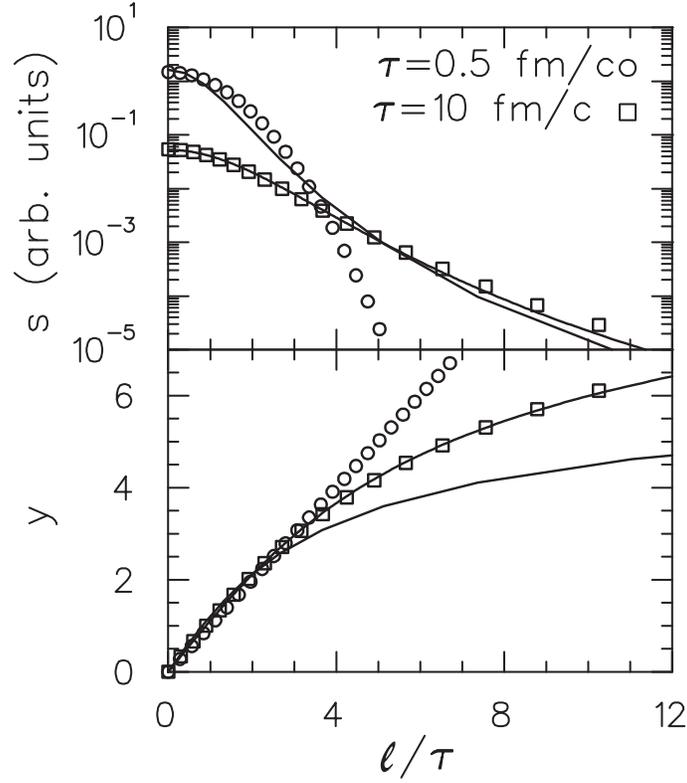}}
\caption{\label{fig:profile_log}
The rapidity and entropy profile are shown as a function of the distance along a hyper-surface of simultaneity. Distances are scaled by the time $\tau$ for easy comparison with a Hwa-Bjorken expasion. The analytic solution (solid lines), which assumed complete stopping at $\tau=0$, is compared to a numerical solution for Hwa-Bjorken initial conditions with $\tau_0=0.5$ fm/c. By $\tau=10$ fm/c (squares), the two solutions are in almost perfect agreement. Even at $\tau=0.5$ fm/c (circles), the two solutions match well for $\ell\lesssim 3\tau$.
}
\end{figure}

Thus, the two solutions are matched by having the two solutions break up at the same reference time along a hyper-surface of simultaneity, and by having the same rapidity spread in the final state $\sigma_y$. Figure \ref{fig:comparison} demonstrates the remarkable agreement of the two evolutions for all but the earliest times by showing $\sigma_y$ and the inverse velocity gradient at the origin as a function of $\tau$. At $\tau=0$, the analytic solution has zero velocity gradient, but by the time the other solution begins at $\tau=0.5$ fm/$c$ has already reached a similar amount of collective velocity. Figure \ref{fig:profile_log} shows that the entire structure of the density and rapidity profiles are very similar at breakup. This shows that the final state of the system tends to move towards the analytic solution at large times, with a given final state being reached by a large number of possible initial conditions. We would expect this trend to be violated for when comparing solutions with very different equations of state.

Although these models have an overly simplistic equation of state and ignore the effects of transverse flow, one can gain some quantitative understanding of the effects of longitudinal expansion at RHIC from the evolutions shown in Fig. \ref{fig:comparison}. First, the variance of the collective rapidity $\sigma_y$ grew by more than 60\% between $\tau=0.5$ fm/$c$ and 10 fm/$c$. Even after incorporating transverse flow or using a softer equation of state, the growth would be several tens of percent. This clearly plays a critical role in understanding the stopping at RHIC, as ignoring the effects of longitudinal acceleration would lead to an underestimate of the initial stopping in central collisions.

The evolution of the inverse velocity gradient at the origin in the lower panel of Fig. \ref{fig:comparison} provides quantitative insight into how longitudinal  acceleration affects the determination of the lifetime from correlations. In correlation analyses, one obtains the inverse velocity gradient by dividing the longitudinal correlation radius, $R_{\rm long}$, by the thermal velocity $v_{\rm therm}$, which is determined from analyses of spectra. Thus, the analyses first determine the inverse velocity gradient, which are identified with the breakup time by assuming an acceleration-less Hwa-Bjorken form for $y$ vs. $\ell$. Fig. \ref{fig:comparison} shows that the neglect of acceleration leads to an underestimate of the break-up time of $\approx 10$\%. As discussed in the previous section, Eq. (\ref{eq:taudistortion}) shows that this underestimate depends only on the ratio of the speed of sound to $\sigma_y$ at breakup. Given that solutions tend towards the analytic solution, this conclusion is somewhat robust. However, lowering the speed of sound or incorporating transverse flow should ameliorate this effect. 

Some analyses of correlation and spectra have been based on assumptions of a purely Boost invariant picture for the longitudinal flow. This leads to an underestimate of the lifetime for two separate reasons. First, the system is not infinite in extent. Convoluting a boost-invariant solution with a Gaussian profile should lead to larger estimates of the lifetime, with the amount of the correction being of the order of 10\%, depending on the $p_t$ range studied. The second effect is the one discussed here, of longitudinal acceleration. Boost-invarian blast waves tend to suggest breakup times near 9 fm/c \cite{Retiere:2003kf}. These times give one pause since the same analyses suggest that the radii have grown from 6 fm initially to 12 or 13 fm, with final collective velocities of 0.7$c$ at the edge. To achieve this final state geometry, the matter would have had to achieve maximum transverse speed immediately, which is unphysical. This inconsistency, which represents part of the ``HBT puzzle'', would be softened by increasing the estimates of the breakup lifetime to 11 or 12 fm/$c$.

\section{One-Dimensional Exponential Solution}
\label{sec:exponential}

Like the first part of the Khalatnikov-Landau solution, we will consider matter ejected from a slab of uniform energy density, thus re-deriving their result but using the mapping procedure applied to the Gaussian profile in the previous section. Unlike the Gaussian solution, all the matter undergoes acceleration in this solution, including the reference point.
We assume a simple form for the rapidity profile and entropy density:
\begin{equation}
y(\tilde{\ell},\tau_R)=y_R(\tau_R)+A(\tau_R)\tilde{\ell},
~~~s(\tilde{\ell},\tau_R)=s_R(\tau_R)e^{-\tilde{\ell}/R(\tau_R)}.
\end{equation}
Here, $\tau_R$ is the time as measured by a reference observer who moves along with the fluid whose local entropy density is $s_R$ and moves with rapidity $y_R$. The distance $\tilde{\ell}$ is the scaled distance measured relative to the reference observer,
\begin{equation}
\tilde{\ell}=\int_{\ell_R}^{\ell} d\ell' \alpha(\ell'),~~~
\alpha(\ell)=\exp\left\{-\int_{\ell_R}^\ell d\ell ~a\right\}=\left(\frac{s}{s_R}\right)^{-c_s^2},
\end{equation}
where $\ell$ is measured along a line of simultaneity, $u\cdot dx=0$. Here, $a$ is again the acceleration of the fluid element. This parameterization will solve the hydrodynamic equations of motion if
\begin{equation}
A(\tau_R)=\frac{1}{\tau_R},~~~R(\tau_R)=c_s\tau_R,~~~s_R(\tau_R)=\frac{B}{\tau_R}
\end{equation}
where $c_s$ is the speed of sound, and $B$ is an arbitrary constant. As was the case with the Gaussian solution $\alpha$ stays fixed for a co-mover in this solution which allows one to state
\begin{equation}
\frac{d\tilde{\ell}}{d\tau_R}=y-y_R.
\end{equation}

Since these solutions are in terms of positions and times relative to  an accelerating observer, expressing the solution in Cartesian space-time coordinates requires first solving for the trajectory of the reference observer. This is not difficult as the observer has velocity $c_s$ at $t=0$ and accelerates with $a=1/\tau_R$ after that point. Since these solutions have the same initial conditions and satisfy the same underlying hydrodynamic equations of motion, they should reproduce the first-stage Khalatnikov-Landau solution. This was checked numerically.

The rapidity distribution can be extracted from the solution by considering $dS/dy$ which should approximately be proportional to the number of particles per unit rapidity. This is only approximate because thermal motion smears out the correspondence between a particles asymptotic rapidity and the rapidity of the source from which it was emitted.  From the solution above,
\begin{equation}
\frac{dS}{dy}=s\frac{d\tilde{\ell}}{dy}\frac{d\ell}{d\tilde{\ell}}=Be^{-(y-y_R)(1/c_s-c_s)},
\end{equation}
where $d\tilde{\ell}/d\ell=\alpha=e^{-c\tilde{x}/\tau_R}$. The precise form of $dS/dy$ depends on the form of the breakup surface in space-time, which is defined by the simultaneity condition here. A remarkable aspect of this result is that the shape is determined solely by the speed of sound, and is independent of the breakup time. Of course, after the shocks meet, matter is no longer being fed into the solution at $y=0$, and the width of the rapidity distribution would no longer be fixed.

\section{Conclusions}
\label{sec:summary}

Here, we have investigated the prospects of solving hydrodynamics in a coordinate system that maintains local simultaneity. This involves storing information about the number and entropy densities and about the velocity gradient along a hyper-surface defined by the condition $u\cdot dx=0$. At each point in time the subsequent evolution of the velocity differences to the next time step become equivalent to those of non-relativistic hydrodynamics. However, maintaining simultaneity requires propagating by a time step $\delta\tau$ that differs at each mesh point by a factor $\alpha=\exp(-\int a\cdot dx)$ relative to a reference observer.

This approach has both advantanges and limitations. One advantage is that the differential equations are linear in the velocity, unlike the usual equations which involve $\gamma$ factors. Since mesh points satisfy $u\cdot dx=0$, it is particularly convenient to extract information about the matter in its rest frame from the mesh. The coordinate system is especially attractive for one-dimensional systems, where for certain cases, one can also scale the position measurement and find some exact analystic solutions by mapping to non-relativistic solutions. 

Working with Simultaneity-conserving coordinates does involve added difficulty when compared to non-relativistic hydrodynamics. First, one must calculate the coefficient $\alpha$ at each mesh point. For ideal hydrodynamics this is not difficult as the coefficient depends only on the local densities, and the densities of the reference observer. For viscous hydrodynamics, calculation of $\alpha$ might involve performing a path integral of the acceleration. Viewing the evolution in standard coordinate systems, e.g., a Cartesian space-time mesh, can also be clumsy as the transformation to a different hyper-surface is a bit cumbersome.

The most serious limitation of the method concerns rotational flow. In the absence of rotational flow the condition $u\cdot dx=0$ defines an infinite hyper-surface in space time, and each step forward in time defines a new surface. After evolving forward in time, each point in space time will belong to one surface, and one surface only. This is not the case for rotational flow, as the condition of simultaneity becomes path dependent. This restriction certainly disqualifies this approach for many applications for relativistic heavy ion collisions.

For certain one-dimensional cases, the method leads to analytic solutions. These are scaling solution, i.e., ones where the density profile is determined by a single parameter. In these cases the hydrodynamic equations of motion exactly reproduce the non-relativistic ones after the distances along the hyper-surface are scaled along with the time. Along with the Khalatnikov-Landau solution for matter ejected from a shock, which was already well known, we were able to find a Gaussian solution. In the scaled coordinates, it involves the expansion of a Gaussian-shaped profile, exactly like the well-known non-relatistic case. One of our biggest disappointments is that the analytic solutions were confined to purely one-dimensional solutions. We could not even find analytic solutions for problems with spherical symmetry. Analytic solutions in higher dimension would have been significantly more useful for testing higher-dimensional relativistic codes.

One virtue of the anlaytic solution was that it provided a transparent means for understanding the effects of longitudinal acceleration at RHIC. The analytic solution, which assumes complete stopping, and a numerical solution of a system with an initial Hwa-Bjorken expansion were compared with the constraint that they acheived the same $\sigma_y$ at breakup. The analytic solution had such significant acceleration, that by a time of 0.5 fm/$c$, when the numerical solution was initialized, it had already achieved a similar amount of collective velocity, and by breakup the two solutions were indistinguishable. For a speed of sound $c_s^2=0.2$, the rapidity spread was shown to increase by over 50\% between $\tau=0.5$ fm/$c$ and a break-up time of 10 fm/$c$. The quickness with which the initial conditions became irrelevant and the numerical solution merged with the analytic one was remarkable, and represents an important lesson. 

The effect of longitudinal acceleration is especially important for determining the lifetime from correlation analyses. The effect of the the expansion was to increase the life time by a simple factor, $2x/D(x)$, where $x$ depended only on the ratio of $\sigma_y$ at breakup to the speed of sound. For RHIC this suggests a 10\% effect. Since the speed of sound and final rapidity spread are well constrained by other measurements, the correction factor factor is quite robust. The main qualifier comes from the fact that the effects of transverse expansion were not coupled into the evolution.

After having investigated prospects in detail, we only enthusiastically endorse simultaneity-conserving approaches for one-dimensional  problems, or for problems for which the symmetry reduces the problem to one dimension. We do not foresee major technical issues for extending the approach to more complicated equations of state or for incorporating viscous effects. For higher dimensions, the difficulties associated with the path integrals and the limitations of ir-rotational flow might bring along more trouble than is warranted by the advantages of the approach.

\section*{Appendix: Other Analytic Solutions}
\label{appendix}

Here, we review three solutions to one-dimensional relativistic hydrodynamics. The first is the Khalatnikov-Landau solution \cite{khalatnikov,landau} which assumes initial conditions of a thin slab with a uniform energy density inside and zero energy density outside. Since no information can travel into the slab, due to the fact that $\partial_xP=0$, a shock develops. In the frame of the discontinuity the matter moves outward from the front at the speed of sound $c_s$. Choosing a coordinate system where the front is at $x=0$, the velocity $v$ and entropy density $s$ outside the slab are described by:
\begin{eqnarray}
v(x,t)=\frac{x+c_st}{t+c_sx},~~~~s(x,t)=s_0e^{-y(x,t)/c_s},
\end{eqnarray}
where $y=tanh^{-1}(v)$ is the rapidity and $s_0$ is the entropy density at $x=0$ which is determined by solving the Rankine-Hugoniot equations and is determined by the equations of state and the energy density inside the slab. This solution requires the additional assumption that the speed of sound is fixed. Once the two shock fronts meet, the solution becomes a real tour de force, and has a much more complicated form than the simple expressions above. The latter part of the solution is expressed in terms of derivatives of an integral of Bessel functions, which can be easily evaluated numerically. 

The second solution is the boost-invariant solution of Hwa and Bjorken \cite{Hwa:1974gn,Bjorken:1982qr} is especially simple and can be easily incorporate an arbitrary equation of state. Contrary to the Landau parameterization, this solution assumes the existence of an initial flow velocity, and in fact, assumes that the flow covers an infinite rapidity window. The solution assumes that all frames which pass through the $z=t=0$ space-time point will see the same evolution. Referred to as ``boost invariance'', this implies that $v=z/t$, and implies zero acceleration. Furthermore, the energy density $\epsilon$, the entropy density $s$, the temperature $T$ and the pressure $P$ depend only on the time measured by co-moving observers, $\tau=\gamma t-\gamma v z=\sqrt{t^2-z^2}$, which is known as the proper time. In the Hwa-Bjorken solution, the distance between neighboring points at the same proper time, as measured by a co-moving observer, is $\delta z=\tau\delta\eta$, where
\begin{equation}
z=\tau\sinh(\eta),~~t=\tau\cosh(\eta).
\end{equation}
Given that $v=z/t$, $\eta$ equals the rapidty $y$ in the Hwa-Bjorken solution. The collective velocity gradient is the same for all co-moving observers, $dv/dx=1/\tau$. 

Current conservation for the boost-invariant solution, $\partial_\mu j^\mu=0$, governs the time evolution of any conserved charge density. Since $\nabla\cdot{\bf v}=1/\tau$, current conservation states that all charge densities fall inversely with time,
\begin{equation}
\rho=\rho_0\frac{\tau_0}{\tau}.
\end{equation}
Since entropy is also conserved, $s$ also falls inversely with $\tau$. Since $P$ and $\epsilon$ are functions of $s$ and the charge densities, their time dependence is determined by the time equation of state. For instance, if the speed of sound is constant, $P=P_0(s/s_0)^{1+c_s^2}$, and $P=P_0(\tau_0/\tau)^{1+c_s^2}$. The boost-invariant nature of the solution can easily incorporate arbitrary equations of state, and can also incorporate viscosity. When viscosity is added, entropy is no longer conserved but one can solve for $s$ as a function of $\tau$ using the Navier-Stokes equation \cite{Gyulassy:1997ib,Cheng:2001dz}.

The Hwa-Bjorken solution would seem justified in the case of extremely high beam energy, where the source covers many units of rapidity. In that limit, the pressure depends only slowly on $\eta$, and since accelerations are driven by pressure gradients, the velocities of particular regions of the matter are fixed, thus justifying the condition $v=z/t$. Boost invariance also seems natural from the perspective of having matter created by the classical fields originating from the ultrarelativistic charges belonging to the incoming hadrons. For an observer at mid-rapidity, the currents appear as $qc$ or $-qc$, where $c$ is the velocity of light. If the observer boosts to a new frame, the currents will still appear to be $qc$ and $-qc$ as long as the the observer does not approach the speed of the colliding hadrons. If the interactions are determined by currents, one would expect an approximate boost invariance at mid-rapidity.

The shortcoming of the Hwa-Bjorken solution comes from ignoring longitudinal acceleration. As we will see in the next section, this shortcoming could result in 10-20\% changes in the interpretation of some observables. One interesting aspect of boost-invariant solutions is that the conservation of total energy is not enforced. Each fluid element expands and does work $PdV$. However, that work never goes into increasing the collective kinetic energy of the fluid elements, since they all coast without acceleration. Instead, the central cell does work on the adjacent cell, which then does even more work on the next cell. For an observer at rest, the work done onto subsequent cells at time $t$ increases exponentially, in a kind of covariant procrastination, until eventually, one reaches the light cone. Again, as long as many units of rapidity are covered, and as long as observations are confined to mid-rapidity, the overall conservation of energy can be ignored.

Recently, Cs\"{o}rg\H{o} et al. \cite{Csorgo:2006ax} published a new set of exact one-dimensional analytic solutions for hydrodynamics. Unlike the Hwa-Bjorken solutions these solutions incorporate longitudinal acceleration. For the Cs\"org\H{o} solution the coordinate $\eta\equiv \tan^{-1}(z/t)$ no longer equals the rapidity. Instead,
\begin{equation}
\label{eq:csorgoassumption}
y=\lambda\eta,
\end{equation}
with $\lambda$ being independent of $\eta$ and $\tau$. A second simplyfing assumption made in \cite{Csorgo:2006ax} is that $P$ depends only on $\tau$. Using Eq. (\ref{eq:csorgoassumption}) one can express the collective velocities and derivatives in the frame of an observer at $\eta_0$ moving with $y=\eta_0$,
\begin{eqnarray}
\partial_\mu=(\partial_\tau,(1/\tau)\partial_\eta),\\
\nonumber
u^\mu=[\cosh(\lambda\eta-\eta_0),\sinh(\lambda\eta-\eta_0)].
\end{eqnarray}
Here, the square brackets are used to represent the two relevant components of the four vector. The divergence of the velocity is then,
\begin{equation}
\partial\cdot u=\frac{\lambda}{\tau}.
\end{equation}
Entropy conservation, $dE=-PdV$, can equivalently be stated in continuous variables as $\partial_\tau\epsilon=-(P+\epsilon)(\partial\cdot u)$, which gives
\begin{equation}\
\label{eq:csorgosconservation}
\frac{\partial}{\partial\tau}\epsilon=-\frac{\lambda}{\tau}(P+\epsilon).
\end{equation}
The equations of motion for hydrodynamics,
\begin{equation}
(u\cdot\partial)u^\mu=\frac{1}{P+\epsilon}\left(\partial^\mu-
u^\mu(u\cdot\partial)\right)P,
\end{equation}
after insertion of the expression for $u^\mu$, and the assumption that $P$ depends only on $\tau$ become,
\begin{equation}
\frac{\lambda}{\tau}[\sinh^2(\gamma\eta),\sinh(\gamma\eta)\cosh(\gamma\eta)]
=\frac{\partial P/\partial\tau}{P+\epsilon}[1,0]
-\frac{\partial P/\partial\tau}{P+\epsilon}[\cosh^2(\gamma\eta),\sinh(\gamma\eta)\cosh(\gamma\eta)],~~~~\gamma\equiv(\lambda-1).
\end{equation}
Both components of the equation will be satisfied either if $\lambda=1$, or 
\begin{equation}
\label{eq:cs1}
\frac{\partial P}{\partial\tau}=-\frac{\lambda}{\tau}(P+\epsilon).
\end{equation}
Dividing this by the similar expression for the energy density, Eq. (\ref{eq:csorgosconservation}), gives the speed of sound
\begin{equation}
c_s^2=\left.\frac{dP}{d\epsilon}\right|_{\rm fixed~S}=1.
\end{equation}
Thus, the Cs\"org\H{o} solution is rather particular, as it only applies when the speed of sound equals that of light. If the dimensionality is $d$, the speed of sound must then be $c_s^2=1/d$ \cite{csorgoprivcomm}.

\section*{Acknowledgments}
The author thanks Kerstin Paech for sharing her insight into hydrodynamics. Support was provided by the U.S. Department of Energy, Grant No. DE-FG02-03ER41259.


\begin{thebibliography}{99}

\bibitem{Huovinen:2003fa}
P.~Huovinen;
\underline{Quark Gluon Plasma 3}, Ed.s: R.C. Hwa and X.N. Wang, World Scientific, Singapore (2003), [arXiv:nucl-th/0305064]
.

\bibitem{Kolb:2003dz}
P.~F. Kolb and U.~W. Heinz;
\underline{Quark Gluon Plasma 3}, Ed.s: R.C. Hwa and X.N. Wang, World Scientific, Singapore (2003);
[nucl-th/0305084].

\bibitem{Teaney:2001av}
  D.~Teaney, J.~Lauret and E.~V.~Shuryak,
  [arXiv:nucl-th/0110037].

\bibitem{Rischke:1996nq}
  D.~H.~Rischke,
  Nucl.\ Phys.\ A {\bf 610}, 88C (1996).

\bibitem{smoothhydro}
R.A. Gingold, and J.J. Monaghan,  Monthly Notices, Royal Astronomical 
Society, Vol. 181, 375-389 (1977).

\bibitem{smoothhydrobook}
G.R. Liu and M.B. Liu, \underline{Smoothed Particle Hydrodynamics,
A Meshfree Particle Method}, World Scientific, Singapore (1996). 

\bibitem{Aguiar:2000hw}
  C.~E.~Aguiar, T.~Kodama, T.~Osada and Y.~Hama,
  J.\ Phys.\ G {\bf 27}, 75 (2001)
  [arXiv:hep-ph/0006239].

\bibitem{landauframe}
L.D. Landau, Izv. Akad. Nauk SSSR 17 (1953) 51.

\bibitem{eckartframe}
C. Eckart, Phys. Rev. 58, 919 (1940).

\bibitem{carterframe}
B. Carter in Relativistic Fluid Dynamics, eds. A. Anile and Y. Choquet-Bruhat (Springer, Berlin, 1989); T.S. Olson and W.A. Hiscock, Phys. Rev. D41 (1990) 3687.

\bibitem{Hwa:1974gn}
  R.~C.~Hwa,
  Phys.\ Rev.\ D {\bf 10}, 2260 (1974).

\bibitem{Bjorken:1982qr}
  J.~D.~Bjorken,
  Phys.\ Rev.\ D {\bf 27}, 140 (1983).

\bibitem{khalatnikov}
I.M. Khalatnikov, Zhur. Eksp. Teor. Fiz. {\bf 27}, 529 (1954).


\bibitem{landau}
S.Z. Belenkij and L.D. Landau, Nuovo Cim. Suppl. {\bf 3S10}, 15 (1956);
{\it Collected papers of L.D. Landaau}, ed. D. ter Haar, Gordon and Breach, New York (1965).

\bibitem{Lisa:2000ip}
  M.~A.~Lisa, U.~W.~Heinz and U.~A.~Wiedemann,
  Phys.\ Lett.\ B {\bf 489}, 287 (2000).

\bibitem{abramowitzstegunpage298}
M. Abramowitz and I.A. Stegun, \underline {Handbook of Mathematical Functions}, Natonal Bureau of Standards, Seventh Printing (1968), p. 298.

\bibitem{Csanad:2003qa}
  M.~Csanad, T.~Csorgo and B.~Lorstad,
  Nucl.\ Phys.\ A {\bf 742}, 80 (2004).

\bibitem{Pratt:1986cc}
  S.~Pratt,
  Phys.\ Rev.\ D {\bf 33}, 1314 (1986).

\bibitem{Makhlin:1987gm}
  A.~N.~Makhlin and Yu.~M.~Sinyukov,
  Z.\ Phys.\ C {\bf 39}, 69 (1988).

\bibitem{Back:2005hs}
  B.~B.~Back {\it et al.}  [PHOBOS Collaboration],hwa
  Phys.\ Rev.\ C {\bf 74}, 021901 (2006).

\bibitem{Renk:2005ws}
  T.~Renk,
  Proc. of Workshop on Particle Correlations and Femtoscopy 2005, Kromeriz, Czech Reupublic, AIP Conf.\ Proc.\  {\bf 828}, 491 (2006)
  [arXiv:hep-ph/0509053].

\bibitem{Retiere:2003kf}
  F. Retiere and M.A. Lisa, 
  Phys. Rev. C70, 044907 (2004).

\bibitem{Gyulassy:1997ib}
  M.~Gyulassy, Y.~Pang and B.~Zhang,
  Nucl.\ Phys.\ A {\bf 626}, 999 (1997).

\bibitem{Cheng:2001dz}
  S.~Cheng {\it et al.},
  Phys.\ Rev.\ C {\bf 65}, 024901 (2002).

\bibitem{Csorgo:2006ax}
  T.~Csorgo, M.~I.~Nagy and M.~Csanad,
  [arXiv:nucl-th/0605070].





\bibitem{Heinz:2004et}
  U.~W.~Heinz and P.~F.~Kolb,
  J.\ Phys.\ G {\bf 30}, S1229 (2004).
  
\bibitem{csorgoprivcomm}
T. Cs\"{o}rg\H{o}, private communication.

\end{thebibliography}
\end{document}